# Reentrant spin-glass and transport behavior of Gd$_4$PtAl, a compound with three sites for Gd

Ram Kumar, Jyoti Sharma, Kartik K Iyer, and E.V. Sampathkumaran*
*Tata Institute of Fundamental Research, Homi Bhabha Road, Colaba, Mumbai 400005, India*

**Abstract**

We report temperature ($T$) dependence (2-300 K) of DC and AC magnetization ($M$), isothermal remnant magnetization ($M_{IRM}$), heat-capacity ($C$), electrical resistivity ($\rho$), and magnetoresistance (MR) of a ternary intermetallic compound, Gd$_4$PtAl, crystallizing in a cubic (space group $F\bar{4}3m$) structure. In this structure, there are three sites for the rare-earth. The magnetization data reveal that, in addition to a magnetic transition at 64 K, there is another magnetic feature below 20 K. The $C(T)$ data reveal an upturn below 64 K, shifting to a lower temperature with increasing field, which establishes that the onset of magnetic order is of an antiferromagnetic type. However, there is no worthwhile feature near 20 K in the $C(T)$ curve. Ac susceptibility peak undergoes an observable change with frequency and, in particular the peak around 20 K gets suppressed with the application of a dc magnetic field; in addition, $M_{IRM}$ undergoes a slow decay with time and isothermal $M$ exhibits low-field hysteresis below 20 K only, which are typical of spin-glasses. The results overall suggest that this compound is a reentrant spin-glass in zero field. There are experimental signatures pointing to the existence of both antiferromagnetic and ferromagnetic components, competing with the variation of temperature and magnetic field, as a result of which electrical and magnetoresistance behaviors are peculiar. The results overall suggest that this compound exhibits interesting magnetic and transport properties.

*Corresponding author. E-mail: sampathev@gmail.com

# 1. Introduction

The rare-earth (R) transition metal (TM) intermetallic compounds with multiple sites for R, are generally interesting from magnetism angle. For example, $Gd_5Si_2Ge_2$ with five symmetrically inequivalent positions remain attractive in the area of giant magnetocaloric effect [1] and $R_7Rh_3$ exhibits [2-5] multiple magnetic transitions as a function of temperature ($T$) and magnetic-field ($H$) with competing ferromagnetic and antiferromagnetic interactions. In this respect, the compounds of the type $R_4$(TM)X type [6], crystallizing in the cubic $Gd_4RhIn$-type structure (in the space group $F\bar{4}3m$) have started attracting some attention [see, Refs. 7-9 and articles cited therein]. In particular, X= Mg, Cd and In compounds have been studied with respect to magnetocaloric effect, cluster spin-glass and hydrogenation behavior. However, for X=Al, only TM=Ru, Rh and Ir based compounds have been known [6]. This prompted Engelbert and Janka [9] to attempt to synthesize materials containing TM= Pd and Pt, and they succeeded in this effort. These authors also reported initial magnetic measurements on some of these compounds, reporting only one magnetic transition temperature for each of these compounds. We however felt it important to carry out a thorough magnetic characterization of such compounds, considering that there are three crystallographically inequivalent positions (24$g$, 24$f$, and 16$e$) for R and magnetism of this family should be much more complicated, judged by the behavior of $R_7Rh_3$ [3-5]. In fact, the authors of Ref. 9 have reported that $Gd_4PtAl$ orders antiferromagnetically ($T_N$= ~64K), despite the fact the sign of paramagnetic Curie temperature ($\theta_p$= ~54 K)) is positive, as though ferromagnetic and antiferromagnetic interactions compete; in support of this, a careful look at the isothermal magnetization ($M$) at 3 K, shown in figure 7 of Ref. 9, shows a metamagnetic anomaly for a small application of $H$, indicating instability of antiferromagnetism. The Ho analogue ordering ferromagnetically below ~24 K indeed exhibits one of the characteristics of spin-glasses, that is, a bifurcation of zero-field-cooled (ZFC) and field-cooled (FC) susceptibility ($\chi$) curves below about 20 K [9], It is therefore worthwhile to study this family in detail. With this motivation, we have undertaken an exhaustive study of magnetic and transport behavior on the compound, $Gd_4PtAl$. We conclude that the magnetism of this compound is more interesting than envisaged earlier [9], mimicking re-entrant spin-glass behavior, with complex magnetization, electrical resistivity ($\rho$) and magnetoresistance (MR) properties. The results are reported in this article.

# 2. Experimental details

The polycrystalline specimen of the intermetallic compound $Gd_4PtAl$ was prepared by arc melting stoichiometric amounts of high purity (>99.9 for Gd; >99.99% for Pt and Al) constituent elements in the presence of argon gas. The molten ingot was found to be single phase within the detection limit (<2%) of x-ray diffraction (XRD) (Cu $K_\alpha$), as confirmed by Rietveld fitting (Fig. 1) to the (cubic) space group $F\bar{4}3m$. [Our attempts to anneal the ingot at 850 C following Ref. 9 are not successful. The reason is that it surprisingly resulted in distortion of the shape of the ingot, as though it melts near this temperature. XRD pattern also confirms that such a specimen is not a single phase]. Therefore we performed further studies on the arc-melted specimen. A scanning electron microscopy (Ultra Field Emission SEM of Zeiss)) was used to make sure that there is no detectable secondary phase and the atomic ratio is uniform (4:1:1) within the detection limit (<2%) of SEM. $T$-dependent DC and AC $\chi$ measurements were performed in different fields on a specimen of 11.6 mg using a commercial (Quantum Design) SQUID magnetometer and the AC field was 1 Oe. Unless specified, all the measurements were performed for the ZFC protocol of the specimen. Isothermal magnetization curves were recorded at several temperatures using the same SQUID. The electrical resistivity was measured as a function of temperature, also in the presence of fixed magnetic fields using a commercial Physical Property Measurements System, PPMS (Quantum Design). A conducting silver paint was used to make electrical contact of the leads with the sample. Isothermal resistivity curves as a function of $H$ at selected temperatures were also collected. The same PPMS was used to measure heat capacity ($C$) on a specimen of 10.8 mg.

# 3. Results and discussions

*3.1. DC magnetic susceptibility*

Fig. 2a shows temperature dependent DC $\chi$, measured in 5 kOe field. Inverse of $\chi$ (shown on the right axis of Fig. 2a) suggests that the Curie-Weiss behavior is obeyed above ~140 K. Below this temperature, there is a deviation from the high temperature linear behavior, which may be due to the gradual development of short range magnetic correlations with decreasing $T$. From the Curie-Weiss fitting of the data above 140 K, the value of paramagnetic Curie temperature ($\theta_P$) is found to be 86 ($\pm$2) K. The positive sign implies that the exchange interaction is predominantly of a ferromagnetic type in this compound. The effective magnetic moment ($\mu_{eff}$) from this fitting is found to be 8.3 ($\pm$0.05) $\mu_B$/Gd$^{3+}$ ion, which is higher than that of the theoretical value of 7.94 $\mu_B$ for Gd$^{3+}$ free ion. It is not unusual to see such an enhanced value for Gd-based intermetallics and this is attributed to conduction electron polarization (possibly involving the 5d band of Pt as well) by the large magnetic moment of the Gd ion. As the temperature is lowered, in the $\chi$ versus $T$ curve, a peak at about 64 K, an upturn below about 40 K, another peak near 5 K, and finally a downturn are observed. Clearly, there are multiple magnetic features in this plot. In order to get a better clarity of the magnetism to avoid misleading conclusions due to the presence of such a large magnitude of field, we measured $\chi(T)$ in a much lower field (100 Oe) for field-cooled (FC) and ZFC protocols, the results of which are shown in Fig. 2b. We observe an additional weak shoulder near 100 K. This peak is apparently smeared when measured in 5 kOe. It is difficult to conclude from the $\chi$ data whether this 100 K feature in the low-field curve is due to the trace of a ferromagnetic impurity (the quantity of which must be below the detection limit of XRD), or intrinsic to the compound. In contrast to Ref. 9, we observe a gradual bifurcation of ZFC and FC curves near 70 K as $T$ is lowered below 100 K. Such an irreversibility of ZFC-FC curves could be easily missed in the event that the remnant field of the magnet of the magnetometer is not nullified before taking ZFC data. The FC curve shows an upturn below about 30 K. This is not anticipated for canonical spin-glasses, but often reported for cluster spin-glass systems in the literature [10-12]. There is a tendency to flatten at a lower temperature (<<10 K). In any case, the bifurcation of ZFC-FC curves is one of the characteristics of spin-glasses. An upturn below 30 K exists even in the ZFC curve, and a peak appears at 20 K. These features with varying temperature suggest complexity associated with the magnetism of this compound. It must however be pointed out that the irreversibility of ZFC-FC curves are sometimes known even in antiferromagnets [13, 14]. We have therefore performed additional studies, described below to explore other spin-glass anomalies.

*3.2. Isothermal remnant magnetization*

We have measured isothermal remnant magnetization ($M_{IRM}$) at selected temperatures, covering different temperature ranges, 2-23 K, 23-64 K, and above 64 K (namely, at 10, 45, 90 and 120 K), as this property has been known to be important to confirm spin-glass behavior [15-18]. For this purpose, the sample was cooled from 150 K (that is, from the paramagnetic state) to the desired temperature, and then a magnetic field of 5 kOe was switched on for 5 min, after which the field was set to zero. It took about 3 mins to attain $H$= 0. $M_{IRM}$ data was collected as a function of time ($t$) after the magnetic field becomes zero. The values of $M_{IRM}$ at $t$=0 (i.e., the time at which $H$ becomes zero) for 10, 45, 90 and 120 K are 0.1805, 0.0107, 0.0005 and 0.00006 emu respectively (with error bars of the order of 0.0002 emu). Clearly, the value obtained in the paramagnetic state is negligibly small as soon as the field is switched off, unlike the situation for 10 K at which the values are significant. The time dependencies of $M_{IRM}$ for 10 and 45 K are shown in Fig. 3 in the logarithmic scale of $t$. It is important to note that the logarithmic plot is not linear and more than one linear region is observed. It is possible that such a slope change is the result of different relaxation behavior of the three types of Gd ions. It is known [15] that the decay of $M_{IRM}$ is quite slow occurring over a period of a few hours for spin-glasses (unlike in ferromagnets or antiferromagnets) and the functional form of the relaxation can be complex varying from one material to the other [15-18]. It is clear from the figure that $M_{IRM}$ decays slowly for 10 K, which is a signature of glassy dynamics. For 45 K, the value of $M_{IRM}(0)$ is much smaller (about 18 times) compared to that at 10 K, and therefore the observed weak change with $t$ can be ignored. Therefore it can be concluded that spin-glass concept is not applicable at this temperature. It is thus safe to assume that, for an intermediate temperature range above 20 K, the spin-glass-like dynamics is lost. This conclusion suggests that this compound can be classified as a re-entrant spin-glass.

*3.3. AC susceptibility*

We have also measured AC $\chi$ as a function of temperature with various frequencies (and an AC field of 1 Oe), in the absence of an external magnetic field as well as in the presence of 5 kOe DC field. The results are presented in Fig. 4. In the absence of a DC field, the real part ($\chi'$) of $\chi$ shows a broad shoulder around 100 K as in the low-field DC $\chi$ measurements, and we are not able to rule out an extrinsic origin of this feature, as stated earlier. This shoulder can be inferred in the imaginary part ($\chi''$) of $\chi$ as well. With the application of 5 kOe field, this feature in $\chi'$ is smeared, as noted in DC $\chi$ data. With further decreasing of temperature, $\chi'$ exhibits a broad peak at around 64 K (at 1.3 Hz) and a peak around 13-20 K, mimicking the behavior of ZFC DC $\chi$ curves measured with 100 Oe. Similar peaks are seen in $\chi''$ curves as well. One can infer from this figure that it is difficult to extract the information about the frequency dependence of the peak - expected for canonical spin-glasses [16, 17] – due to the large widths of the peaks, both in $\chi'$ and $\chi''$. Hence, it is not meaningful to carry out any further quantitative analysis. It is however noted that the values undergo a marginal decrease with increasing frequency below 60 K, as though there is a frequency dependence of the peaks. It may be added that the $\chi''$ curves measured with 133 and 1333 Hz are so noisy that we are not able to show these data. In $\chi'$, in the presence of 5 kOe field, the 13-20K-feature is completely suppressed; however, the peak around 60 K still persists. This is consistent [16, 17] with the conclusion that 60K-peak does not arise from spin-glass freezing, while 13-20 K peak (in zero field) originates from spin-glass freezing. This inference is consistent with the behavior of $M_{IRM}(t)$. The in-field $\chi''$ curves are featureless, which is consistent with spin-glass dynamics in zero-field. Finally, $\chi'(T)$ shows up a peak near 35 K in 5 kOe, which may be compared with the appearance of an upturn in DC $\chi$ in 5 kOe (Fig. 2a). This is baffling us and could be an artifact of differences in the field-induced magnetization changes at low temperatures or of indication of subtlety of spin-dynamics.

*3.4. Heat capacity*

Fig. 5 shows the heat capacity as a function of temperature (2-120 K) for different fields of 0, 10 and 50 kOe. $C$ decreases with decreasing $T$ as expected due to dominating lattice contribution down to about 70 K, below which there is an upturn resulting in a $\lambda$-like anomaly. This establishes the existence of a long-range magnetic transition around this temperature (also see the inset of Fig. 5). We also measured $C(T)$ in the presence of external magnetic fields and the peak (occurring at 64 K) gets shifted to a lower temperature, say, in 10 kOe. This supports that the 64K-feature is due to antiferromagnetic ordering. This is interesting considering that the sign of $\theta_p$ is positive, which suggests ferromagnetic ordering should be favoured. However, there is no notable peak at lower temperatures (say, near or below 20 K) or near 100 K. The absence of any feature near 100 K appears to favor different explanations for the 100K-magnetic feature presented above, that is, either in terms of magnetic impurities or in terms a weak magnetic feature involving low entropy. Since we are not able to convincingly understand this feature, we leave this question open for future investigators. Though a careful look at the $C/T$ curve reveals a very weak shoulder (or a change of slope near 18 K), a monotonic decrease of $C$, in other words, the absence of a prominent feature well below 64 K – in particular below 25 K – is consistent with a spin-glass-like transition at such low temperatures. It is to be noted that the data below 10 K can be fitted to the $T^{3/2}$ form, which supports dominance of ferromagnetic correlations in this magnetic phase. In short, $C(T)$ provides evidence for the existence of antiferromagnetic order at about 64 K, transforming to a spin-glass phase at lower temperatures with a dominant ferromagnetic exchange interaction. We could not derive any information about magnetic part of $C$ or magnetic entropy due to the lack of a reference for lattice contribution, as our attempts to synthesize non-magnetic counterparts failed.

*3.5. Isothermal DC magnetization and isothermal entropy change*

We have recorded isothermal magnetization at several temperatures (in a few degree interval), not only for a better understanding of the changes in the magnetic character with $T$, but also to throw light on the magnetocaloric behavior. We first focus on the low-field hysteresis curves at selected temperatures, shown in Fig. 6 for the field variation 0→10→-10→0→10 kOe. It is transparent that the plots below 20 K are hysteretic, unlike the ones for $T>20$ K. This is supportive of the existence of ferromagnetic component below $T=20$ K. A careful look at this figure however reveals that, in the low temperature range ($T \leq 20$ K) for the virgin curve, there is an increase of slope beyond about a few kOe, which is a characteristic of antiferromagnets exhibiting a spin-reorientation. The fact that the virgin curve lies outside the envelope curve for $T<20$ K establishes the existence of a disorder-broadened metamagnetic transition [19], which can happen in the event that an antiferromagnetic component exists. Thus, the ferromagnetic and antiferromagnetic signatures coexisting in these plots offer support to the spin-glass nature of magnetism below 25 K. Absence of hysteresis is noted for $T$ above 20 K (see the plots of 35 and 60 K), as though spin-glass freezing is absent at such higher temperatures; it however appears that there is a sharp, but weak, increase initially, followed by a curved regime, supporting that the compound contains a ferromagnetic component superimposed over an antiferromagnetic matrix. Even for 80 and 105 K, such a weak increase at low fields superimposed over linear region is noted. This signals the existence of a weak ferromagnetic component around 100 K in a matrix of paramagnet.

Looking at the high-field behavior (measured for the virgin state only up to 70 kOe), $M(H)$ curves (Fig. 7a) do not show the tendency to saturate even in 70 kOe at all temperatures in the magnetically ordered state - an observation which rules out exclusive ferromagnetism implied by the positive sign of $\theta_p$. We have extrapolated the linear region above 55 kOe in the $M(H)$ plots to zero-field and the extrapolated saturation magnetization, $M_{sat}$, thus obtained (shown in the inset of Fig. 7a) exhibits a concave curvature at the onset of magnetic order, rather than the sharp change in the event of ferromagnetic order; the observed $M_s$ values, even at 2 K, is far less (<4 $\mu_B$ per Gd) than that expected for Gd ion (7 $\mu_B$), implying the role of an antiferromagnetic component. Finally, it is to be remarked that non-zero values of $M_{sat}$ well above 100 K, can be seen even at temperatures as high as 150 K. This finding supports the existence of short range ferromagnetic order extending up to about 150 K, which is consistent with the inference from the inverse $\chi$ plot.

We have also inferred the behavior of magnetocaloric effect from the values of isothermal entropy change ($\Delta S$) (shown in Fig. 7b) from these $M$-$H$ curves, using the Maxwell relation [20]. The curves shown are for a change of field from zero to 20, 50 and 70 kOe. $-\Delta S$ lies in the positive quadrant, supportive of a dominant ferromagnetic component when the magnetic field is applied [20]. There are peaks near 68 K and 25 K, consistent with the fact that there are magnetic features involving entropy changes around these temperatures. $-\Delta S$ attains a reasonably large value of about 8.5 J/kg K at the peak (at around 68 K) for the final field of 70 kOe. The peak-value of $-\Delta S$ around 25 K is smaller than that obtained at 68 K, which implies competing magnetic interactions. It is to be noted that the values above 68 K decay very slowly till about 150 K with the sign of $-\Delta S$ remaining positive around 100 K. This finding appears to support the proposal that ferromagnetic clusters gradually form well above $T_N$.

*3.6. Electrical resistivity and magnetoresistance*

Fig. 8 shows electrical resistivity as a function of temperature, measured in the presence of 0, 10, 30, 50, 100 and 130 kOe fields. In zero field, the derivative, $d\rho(T)/dT$, is positive endorsing metallicity of the material. What is interesting is that the zero-field and in-field curves do not merge, even at room temperature, and that the values are higher in the presence of an external $H$ in the paramagnetic state. In other words, the sign of MR, defined as $[\rho(H)-\rho(0)]/\rho(0)$, is positive in the paramagnetic state. It is often reported [21] that the negative MR component from the paramagnetic Gd is large compared to the metallic contribution from the conduction electrons. Hence MR is generally negative for such Gd systems in the paramagnetic state, even in those containing relatively less atomic percent of Gd. It is therefore interesting that positive MR is seen in the paramagnetic state, despite having about 66 atomic percent of Gd. Clearly, this compound presents an interesting situation in which conduction electron contribution to scattering process dominates the one due to Gd 4f electrons. In the zero field curve, below 100 K, there is an upturn, which should not be in principle seen in the paramagnetic state of Gd systems. But such upturn can in some cases arise from the bulk of the sample (and not due to impurities) have been established by us in the past [21], the exact origin of which is yet to be resolved. It was proposed that building of short-range ferromagnetic clusters before the onset of long-range magnetic order (called "magnetic precursor effect") in some fashion is responsible for this. It is at present not clear whether it is due to magnetic superzone formation of the magnetic clusters before long-range ordering sets in. Following another magnetic transition near 64 K, $\rho$ shows a small dip; below 40 K, there is again a sharp upturn, with a peak around 17 K and then a fall due to possible onset of spin-glass phase. In fact, there is an extremely weak upturn at further lower temperatures (below 4 K). Clearly, the electronic scattering behavior of this compound in the magnetically ordered state as a function of temperature is quite complex due to competing magnetic phases with magnetic Brillouin-zone boundary gaps [22]. (For this reason, the residual resistivity ratio can not be taken as a measure of disorder in this material). It is of interest to focus future studies to understand this aspect better. With the increase of magnetic field, it appears that the 17K-peak gets shifted gradually towards the higher temperature range, indicating dominance of ferromagnetic component in the spin-glass phase (which is also inferred from the $T^{1.5}$ form of $C$, mentioned above). As a result of this upward shift of this peak, the distinct feature that occurs in zero field around 64 K is apparently getting suppressed in high fields. The negative slopes in the plots below 64 K, though appear robust to small applications of field, are gradually diminishing with $H$ and these are absent, say for $H$> 100 kOe. These results suggest that the magnetic gaps are getting closed gradually with increasing $H$. It is however fascinating that the 100K-minimum in $\rho(T)$ persists even at high fields, but gradually shifting towards higher temperatures, for instance, to about 140 K in 130 kOe. This upward shift – clearly resolved in the resistivity data - signals that the 100K-minimum arises from ferromagnetic clusters. It is for this reason that we are not able to rule out completely the role of such intrinsic clusters on the 100K-magnetic feature. The reason is that, as far as electrical transport process is concerned, a contribution by an impurity is expected to get overshadowed by that of the majority phase. Thus, viewed together with the results from other techniques, these findings, suggest the existence of a series of features, at ~100 K (a minimum), ~64 K (due to antiferromagnetism), and ~20 K (due to spin-glass), possibly due to the $T$-dependent complex exchange interactions from the Gd ions at the three sites.

In order to get a better picture of MR behavior, we have measured ρ as a function of field at selected temperatures and the results are shown in Fig. 9 in the form of MR versus $H$. Let us look at the curves in three different ranges: (i) Below 20 K: Barring initial insensitivity to $H$ (up to about 5 kOe), MR remains in the negative quadrant for <15 K; for 15 K and 20 K - that is, when the compound is on the verge of entering nearby antiferromagnetic phase with increasing $T$ - positive MR tends to contribute, as a result of which the values are small and positive for initial fields. Ignoring the initial small positive values, the negative sign of MR in general for T<20 K is consistent with spin-glass freezing (due to the gradual orientation of spins by external magnetic field). (ii) 20 K<$T$<70 K: In this region, MR is distinctly positive and significant in its value, undergoing quadratic variation with $H$ initially (see, for instance, the initial curvature for 30 and 50 K); this is consistent with antiferromagnetism [23] adding to (classical cyclotron motion induced) metallic contribution [24]; however, with increasing fields beyond a certain value (e.g., 20 kOe for 30 K, 30 kOe for 50 K), the ferromagnetic alignment by the external field reduces electronic scattering, thereby causing a downward trend of MR; at very high fields, say around 100 kOe, for 30 K, the ferromagnetic alignment is large enough to make the sign of MR negative (due to the suppression of residual spin-fluctuations). (iii) $T$ >64 K: The curves are in the positive quadrant only without any tendency for sign reversal; the MR varies quadratically with $H$; this means that the conduction electron contribution dominates in this temperature range.

Finally, we remark that a low-field hysteresis is observed in MR versus $H$ plots up to 20 K, which is in good agreement with that seen in corresponding $M$-$H$ loops (see, for instance, the inset of Fig. 9 for the behavior at 1.8 K).

## 4. Summary

The results reported in this article on a ternary intermetallic compound, Gd$_4$PtAl, reveal that this compound is a reentrant spin-glass, with magnetic order setting in near 64 K and spin-glass features around 20 K. Since there are strong indications in our results for the delicate balance between ferromagnetism and antiferromagnetism, spin-glass anomalies may be associated with the competition between these two interactions, and this may be in some way related to the complexity associated with the exchange interaction among Gd ions at three different sites. Possible role of crystallographic order as a source of spin-glass features is not clear at this moment. There is another magnetic feature around 100 K; though there is some indication from transport data that this could be intrinsic, we can not exclude the role of trace of a magnetic impurity (not detectable by XRD and SEM). In support of re-entrant behavior, the $C(T)$ data reveal an upturn below 70 K, shifting to a lower temperature with increasing field, which confirms antiferromagnetism setting in at 64 K; however, there is no worthwhile feature near 20 K in $C(T)$. $M_{IRM}$ undergoes a slow decay with time and isothermal $M$ exhibits low-field hysteresis below ~20 K only, which are typical of spin-glasses.

The temperature coefficient of ρ is positive at higher temperatures (>100 K) in the paramagnetic state, but it shows peaks and negative slopes with a complex $T$-dependence as one enters glassy phase.

These suggest the existence of magnetic gaps in the magnetically ordered state. It is interesting that there is a minimum in $\rho(T)$ around 100 K.

Isothermal $M$ in the magnetically ordered state, after a rise for initial applications of field, keeps rising with magnetic field without saturation at high fields (measured up to 70 kOe), indicating the existence of antiferromagnetic component at all temperatures in the magnetically ordered state. However, positive paramagnetic Curie temperature, isothermal entropy change, functional form of low-temperature heat-capacity and the response of the low-temperature $\rho(T)$ peak to the application of $H$ reveal the existence of a ferromagnetic component.

Additional features to be noted are: (i) In the low-field hysteresis plots of $M(H)$ below 25 K, the virgin curve lies outside the envelope curve, as though there is a first-order magnetic transition for a small application of $H$, as known as $Nd_7Rh_3$ [4]. (ii) The magnetoresistance, defined as MR= $[\rho(H)-\rho(0)]/\rho(0)$, is positive not only in the antiferromagnetic state for reasonably large applications of magnetic field, but also in the paramagnetic state, as though metallic contribution to MR dominates; however, in the spin-glass phase, as expected, MR remains negative.

The results overall suggest that this compound exhibits interesting magnetic and transport behavior as a function of temperature and magnetic field. It is of interest to focus future work to understand the role of three sites and mutual interactions between these sites on these properties.

Acknowledgements: This work was supported by SERB JC Bose Fellowship Grant Number SR/S2/JCB-23/2007 dated 16/8/2018.

**Figure 1:** X-ray diffraction pattern of $Gd_4PtAl$. The continuous line through the data points is the result of Rietveld refinement. The difference between the experimental pattern and the line obtained by fitting are shown at the bottom (in blue). The fitting parameters (including R-factors) including χ (which here represents goodness of fit) are also shown.

**Figure 2:** Magnetic susceptibility as a function of temperature (2-300 K) measured in (a) 5 kOe, and (b) 100 Oe for $Gd_4PtAl$, while warming. In (a), inverse susceptibility (right axis) is also plotted, with the line representing Curie-Weiss fitting to the data in the high-temperature region. In (b), the data for zero-field-cooled (ZFC) and field-cooled (FC) conditions are shown and the lines drawn are guides to the eyes.

**Figure 3:** Isothermal remnant magnetization as a function of time ($t$) in logarithmic scale, measured as described in the text, for 10 and 45 K for $Gd_4PtAl$. The values are normalized to respective $t= 0$ data point and $t= 0$ is defined as the time immediately after the applied magnetic field becomes zero.

**Figure 4:** Real part (top) and imaginary (bottom) parts of ac susceptibility as a function of temperature, measured with different frequencies in the absence and in the presence of external DC magnetic field (5 kOe), for $Gd_4PtAl$.

**Figure 5:** Heat-capacity as a function of temperature in zero-field, 10 and 50 kOe for $Gd_4PtAl$. Inset shows the plot of heat-capacity divided by temperature in zero-field and vertical arrow shows where there is a feature due to antiferromagnetic ordering.

**Figure 6:** Magnetic hysteresis loops at various temperatures for $Gd_4PtAl$. For 105 K, a dotted line is drawn through the data points in the range 6 to 10 kOe.

**Figure 7:** (a) Isothermal magnetization behavior at selected temperatures for $Gd_4PtAl$. Inset shows the saturation magnetization per Gd, obtained by linear extrapolation of the data above 55 kOe to zero-field. (b) Isothermal entropy change as a function of temperature for a variation of the magnetic field from zero to a desired field, derived from isothermal magnetization data. The lines through the data points serve as guides to the eyes.

**Figure 8:** Temperature dependence of electrical resistivity in the presence of several fields for $Gd_4PtAl$.

**Figure 9:** Isothermal magnetoresistance at selected temperatures for $Gd_4PtAl$ for the zero-field-cooled condition of the specimen (for increasing field direction only). In the bottom plot, the curves below 20 kOe only are shown to highlight the low-field behavior. In the inset, hysteretic behavior at 1.8 K is shown.

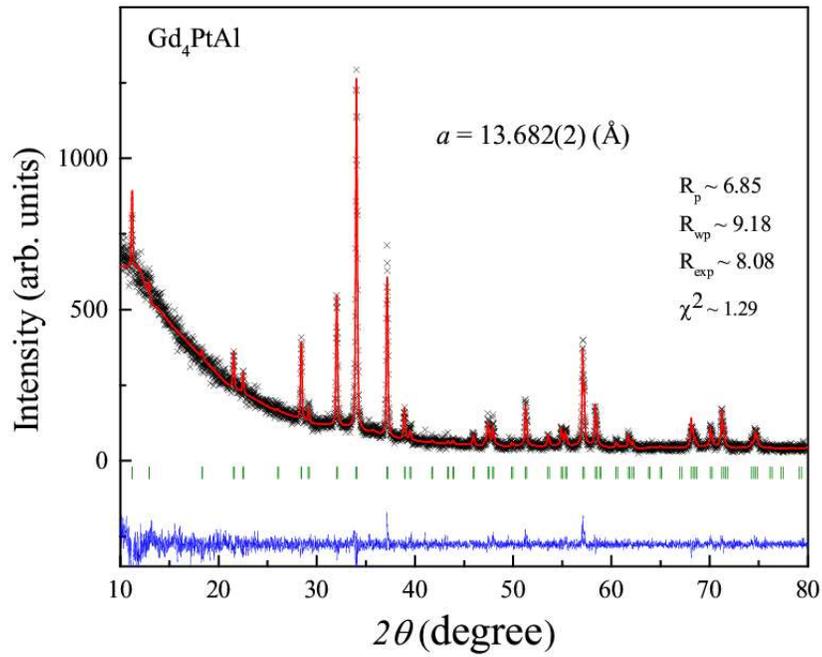

Figure 1

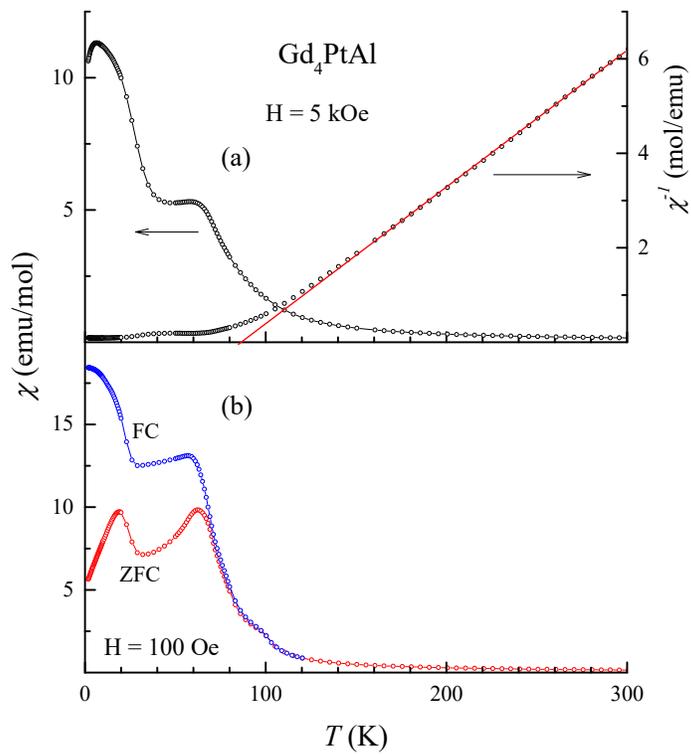

Figure 2

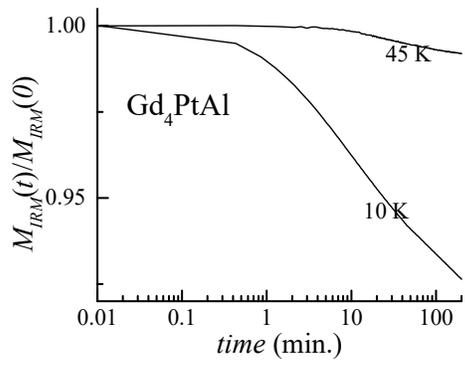

Figure 3

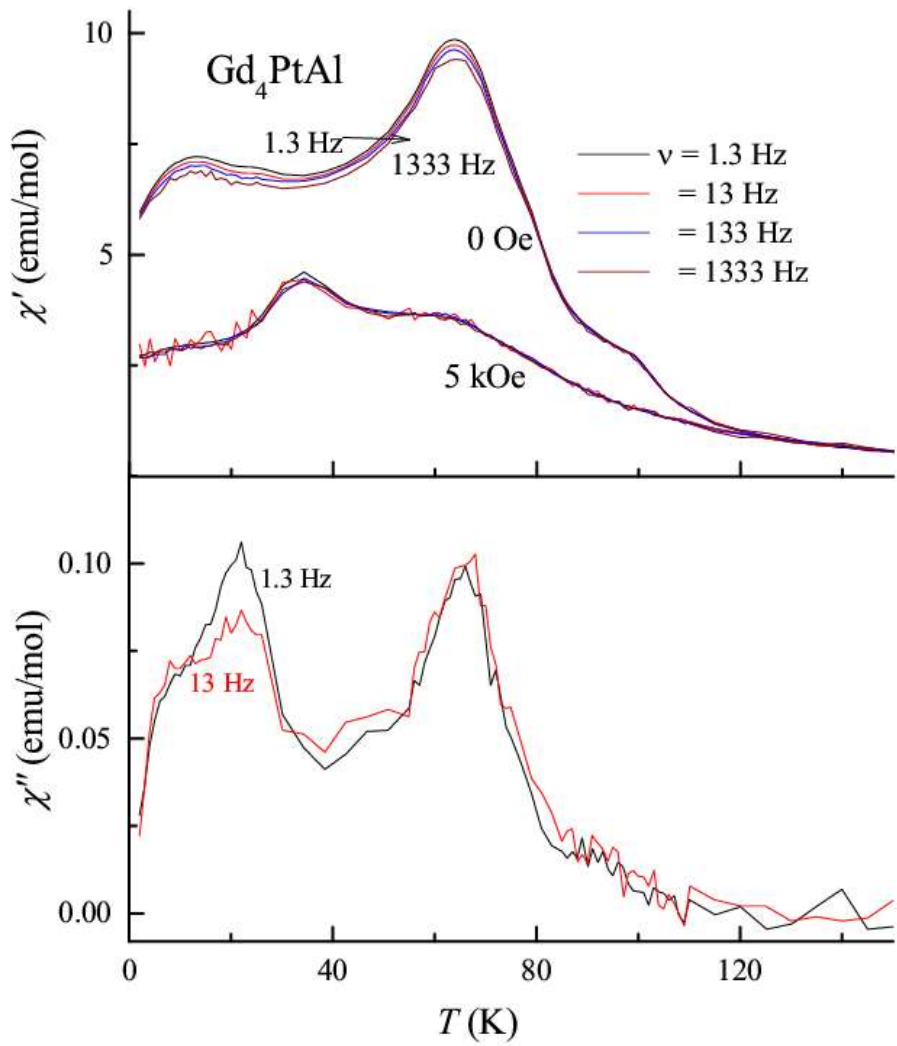

Figure 4

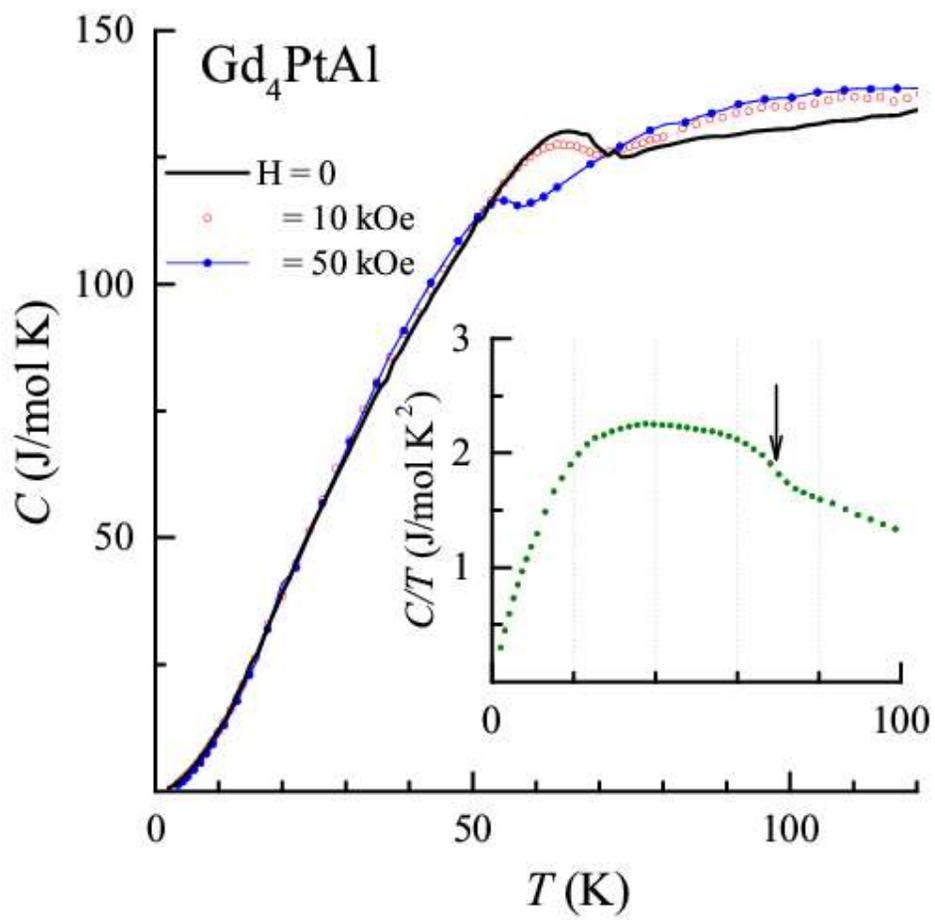

Figure 5

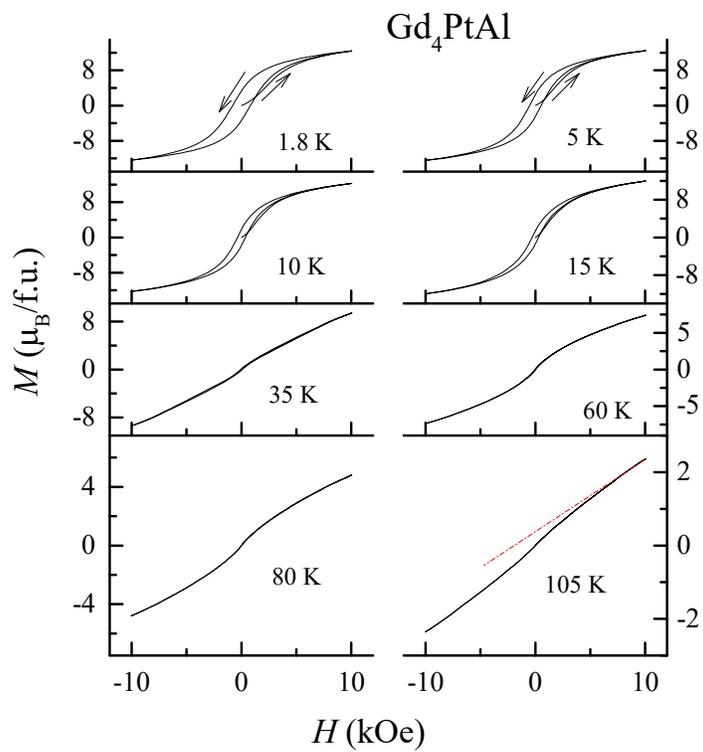

Figure 6

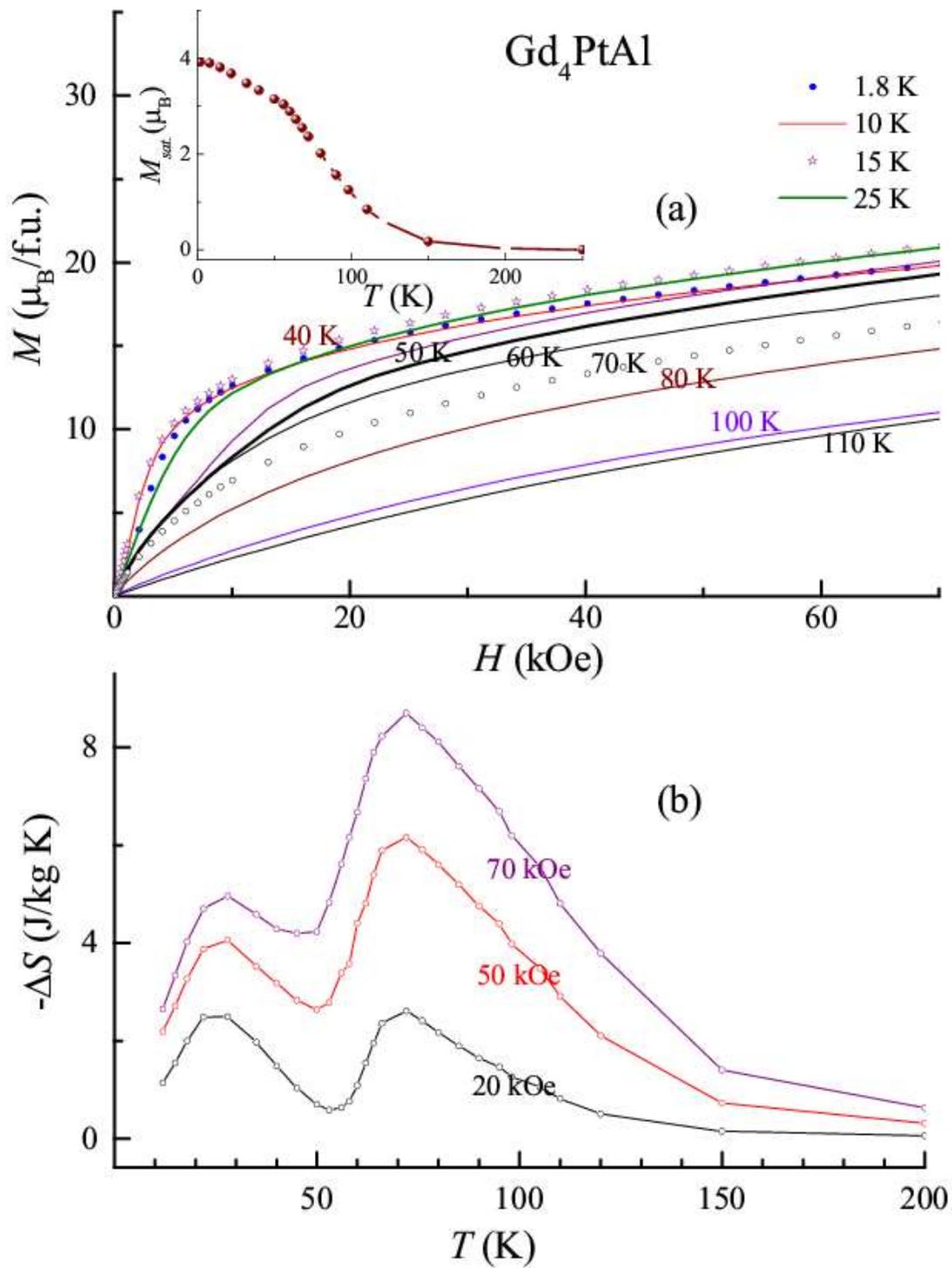

Figure 7

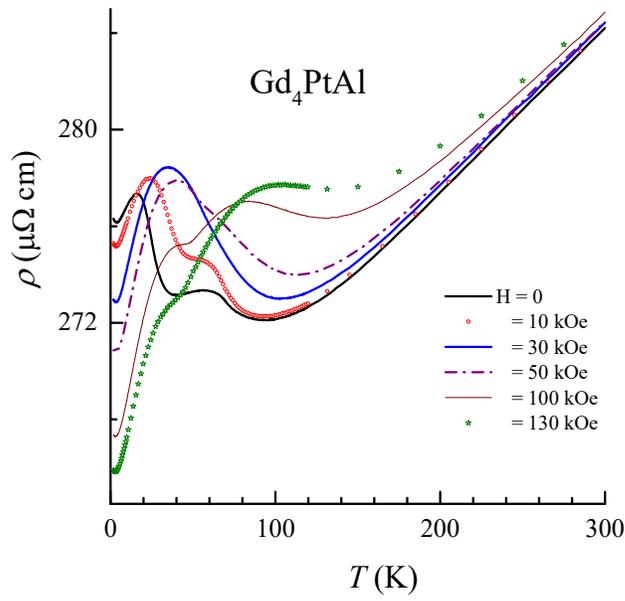

Figure 8

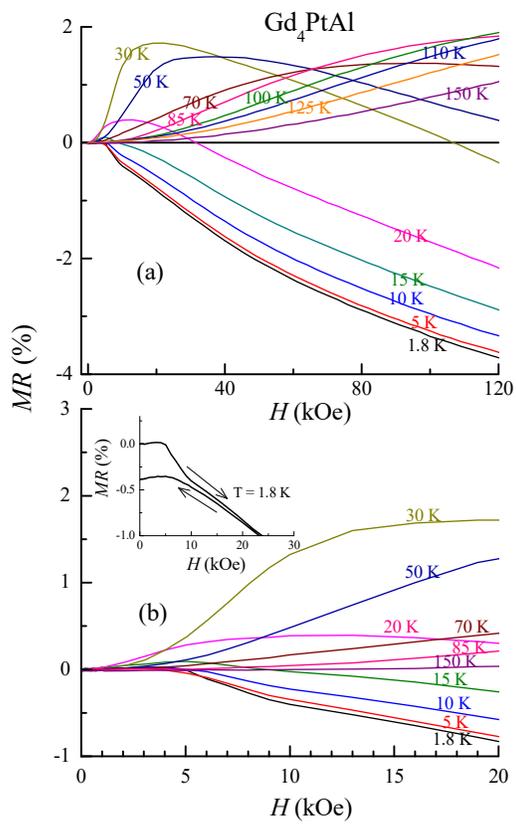

Figure 9